\documentclass[num-refs]{wiley-article}
\usepackage{siunitx}
\usepackage{amsmath}
\usepackage{graphicx}
\usepackage{tabularx}
\usepackage{epstopdf}
\usepackage{color}
\usepackage{hyperref}
\usepackage{dcolumn}
\usepackage{bm}
\usepackage{ulem}

\papertype{Full Paper}
\paperfield{Submitted to Magnetic Resonance in Medicine}

\title{Improving $B_{1}$ homogeneity in abdominal imaging at 3 T with light and compact metasurface}

\author[1]{Vsevolod~Vorobyev}
\author[1]{Alena~Shchelokova}
\author[1,2]{Alexander~Efimtcev}
\author[3]{Juan~D.~Baena}
\author[4]{Redha~Abdeddaim}
\author[1]{Pavel~Belov}
\author[1]{Irina~Melchakova}
\author[1]{Stanislav~Glybovski}

\affil[1]{Department of Physics and Engineering, ITMO University, Saint Petersburg, 197101, Russian Federation}
\affil[2]{Department of Radiology, Federal Almazov North-West Medical Research Center, Saint Petersburg, 197341, Russian Federation}
\affil[3]{Department of Physics, Universidad Nacional de Colombia, Bogota, 111321, Colombia}
\affil[4]{Aix Marseille University, CNRS, Centrale Marseille, Institut Fresnel, F-13013, Marseille, France}

\corraddress{Alena~Shchelokova, Department of Physics and Engineering, ITMO University, Saint Petersburg, 197101, Russia}
\corremail{a.schelokova@metalab.ifmo.ru}

\runningauthor{Vsevolod Vorobyev et al.}

\begin{document}

\maketitle

\begin{abstract} %max 250 words!
\small
\textbf{Purpose}:
Radiofrequency field inhomogeneity is a significant issue in imaging large fields of view in high- and ultrahigh-field MRI. Passive shimming with coupled coils or dielectric pads is the most common approach at 3 T. We introduce and test light and compact metasurface, providing the same homogeneity improvement in clinical abdominal imaging at 3 T as a conventional dielectric pad.
\\
\textbf{Methods}: The metasurface comprising a periodic structure of copper strips and parallel-plate capacitive elements printed on a flexible polyimide substrate supports propagation of slow electromagnetic waves similar to a high-permittivity slab. We compare the metasurface operating inside a transmit body birdcage coil to the state-of-the-art pad by numerical simulations and \textit{in vivo} study on healthy volunteers.
\\
\textbf{Results}: 
Numerical simulations with different body models show that the local minimum of $B_1^+$ causing a dark void in the abdominal domain is removed by the metasurface with comparable resulting homogeneity as for the pad without noticeable SAR change. \textit{In vivo} results confirm similar homogeneity improvement and demonstrate the stability to body mass index.
\\
\textbf{Conclusion}:
The light, flexible, and cheap metasurface can replace a relatively heavy and expensive pad based on the aqueous suspension of barium titanate in abdominal imaging at 3 T.

% Please include a maximum of seven keywords
\keywords{abdominal imaging, 3 T, $B_{1}$ inhomogeneity, dielectric pad,  metasurface, image shading, passive shimming}

\end{abstract}

\newpage
\section{Introduction}
Radiofrequency (RF) field inhomogeneity is a significant concern in high-field and ultrahigh-field imaging of a large Field Of View (FOV). It starts to appear at 3 T in abdominal imaging and becomes even more pronounced at higher fields (e.g., body and brain imaging at 7 T). At 3 T, the RF wavelength measures $\approx 240$ cm in the air while getting shortened to $\approx 26$ cm when measured inside the body tissues. In this case, such wave effects as phase delay and reflection occur in the abdominal domain causing local constructive or destructive interference of $B_1^{+}$ field excited by any volume coil. This leads to corresponding brighter or darker areas of the image signal, i.e., the artifact referred to as the dielectric resonance, standing wave, or RF interference. At 3 T, the transverse FOV dimensions in most body imaging studies, such as abdominal and spinal imaging, and imaging of large water volume (e.g., pregnancy, ascites), are comparable to the wavelength. Therefore, the dielectric resonance effect highly deteriorates the imaging quality \cite{Christianson2012DukeRO}.

Clinical studies of the abdominal organs are associated with many nuances, the combination of which can play against obtaining high-quality diagnostic information. So, for individuals with large Body Mass Indices (BMIs), the artifact causes a decrease, or even loss of signal, in the area of the right and/or left lobe of the liver, the abdomen middle floor, and the pancreas \cite{yang2010optimizing}. The resulting signal minimum is pronounced when using fast spin-echo (FSE) and true fast imaging with steady-state precession (trueFISP) pulse sequences, which are the most important in assessing pathological changes in any localization. The radiologist relies on them to describe and clarify the nature of changes in images with different contrast. The decision to use additional protocols or contrast enhancement is often made during the scan, based on the T2-weighted images (T2WI). For example, small pancreatic neoplasms are sometimes asymptomatic and are found by accident. Therefore, in the presence of a pathological process in the artifact zone, the pathology may be missed, and the patient will not receive the necessary treatment. Tuning the parameters of these pulse sequences helps to avoid signal loss in the affected zone to a certain extent. The desired effect may be achieved by increasing the gap between slices or using spin-echo (SE) pulse sequences. However, in the first case, this will negatively affect the detection of small pathological changes. In the second one, it is inapplicable in the study of the abdomen since breath-hold is required.  Moreover, the above artifact is of particular relevance in the study of the fetus or placenta when the data obtained cannot be adequately interpreted, and the information content of the study drops significantly~\cite{Semenova2020}. Definitely, in the above-described cases, additional measures to remove the artifact and improve the homogeneity of $B_1^{+}$ at the hardware level become unavoidable.

Various methods have been developed to homogenize the $B_1^{+}$ level across the Region Of Interest (ROI), including the parallel transmit (pTx) and passive shimming by coupled coils or dielectric pads (DPs). The pTx approach~\cite{Grissom_replace_1,B_grissom2010,C_Raaijmakers2011} carries out RF shimming by using multi-element transmit coils, elements of which are individually driven simultaneously with custom RF pulses sharing a standard gradient waveform. While the pTx approach has become useful for biomedical research, its application in clinics is limited by the hardware complexity and difficulties in ensuring a patient's RF safety during RF excitation with different shimming scenarios.
The first passive shimming method uses surface RF coils \cite{schmitt2005b1, Wang2009}, passively coupled to a driven volume transmit coil. The coupled coil typically has $\approx10$\% higher resonant frequency than the transmit coil. Therefore, the transmit coil induces conduction current in the coupled coil with a proper phase so that the corresponding secondary $B_1^{+}$ field adds constructively to the primary transmit field in the artifact region. However, the coupled coil has a considerable weight decreasing the patient's comfort, and its presence between the human body and a local receive coil may decrease the latter's sensitivity. Recently, an arrangement of four resonant wires has been proposed as a coupled element which possesses symmetric (electric) and anti-symmetric (magnetic) type of the induced current oscillations which facilitates the field pattern improvement without undesirable feedback to the transmit coil \cite{RedhaPRX}.   
The second widely known passive shimming approach is based on the application of high-permittivity dielectrics. Many examples of DPs locally modifying $B_1^{+}$ field of a standard transmit body birdcage coil (BC) have been shown in the literature for cardiac  \cite{brink2014high, de2016improved, brink2015effect}, abdominal \cite{de2012increasing}, and fetal \cite{van2019simulation} imaging at 3 T. In contrast to the passive coils, in this method, the secondary field required for homogeneity improvement is created due to induced displacement currents. Thus, by placing a DP over the ROI, the overall $B_1^{+}$ level is equalized \cite{Webb2011}. The thinner the pad, the higher its relative permittivity is required to remove the artifact \cite{Teeuwisse2012}. Pads should be specially designed for different ROIs and vary in dimensions, thickness, and permittivity of the filling material, usually based on mixtures of liquids and ceramic powders. Despite being a widely-accepted and simple technique, composite ceramic pads have several common drawbacks. Their material parameters can change with time, some materials they may contain are bioincompatible, and they are heavy (typically weight several kilograms) \cite{Neves2017}.

One of the proposed ways of solving DPs' problems is replacing the composite ceramic-based mixture with an artificial dielectric metamaterial. The previous study \cite{vorobyev2020artificial} has reported that for brain imaging at 7 T, a DP can be replaced by an \textit{artificial dielectric slab} made of stacked Printed Circuit Boards (PCBs) carrying periodically arranged copper patches. The signal homogeneity was shown to improve similarly for the pad and slab when their dimensions are the same, and the effective permittivity of the artificial dielectric is equal to the pad ones. The proposed slab consisted of low-cost boards, its properties were stable in time, and it was much lighter in weight than the DP. However, the proposed slab was rigid and operated in only one position to the head (parallel to the axial plane). Therefore, it allowed homogenizing $B_1^{+}$ field only in the parietal lobes, whereas temporal lobes imaging is more important in clinical studies. Therefore, for brain imaging at 7 T, a PCB structure is needed to replace the DP in a position parallel to the sagittal plane. Moreover, it is rather important to replace DPs at 3 T, where the dedicated DPs for cardiac and abdominal studies, typically placed in parallel to the coronal plane, are especially heavy.

In this paper, we introduce a better alternative to the DPs. An ultra-thin and light periodic structure called \textit{the metasurface (MS)} is proposed for removing the $B_1^{+}$ inhomogeneity artifact in abdominal imaging at 3 T, providing the same effect as a dedicated DP. 

MSs are thin two-dimensional periodic structures with periodicity and thickness both much smaller than the wavelength, which can modify the applied electromagnetic field in a desirable way \cite{GLYBOVSKI20161}. When excited by the external field, each unit cell of an MS produces a secondary RF field. The secondary field collectively created by all periodically arranged unit cells of the MS is similar to the field of a uniform sheet with electric and/or magnetic surface currents. By engineering the unit cells, it is possible to control the RF field's characteristics, such as magnitude, phase, and polarization, at least at distances larger than one period from the MS. MSs were first introduced in MRI in \cite{LeshaAdvMater}, where a periodic rectangular structure of thin parallel wires immersed in water was used as a resonator passively coupled to a BC. This MS almost entirely concentrated the whole energy of $B_1^{+}$ field of the coil in the wires' vicinity due to resonant excitation of one of the resonant surface modes. This effect locally increased the transmit efficiency and Signal-to-Noise Ratio (SNR) of the BC. The water-filled wire MS was followed by several modifications proposed for the same type of resonant focusing operation~\cite{GlybovskiRadio,LeshaSciRep,AlenaPRAppl,SHCHELOKOVA201878,AlenaMRM,Brui,Kretov,Saha2020}, and some alternative MS-based coupled resonators were proposed (e.g.,~\cite{Yang,Issa}).

Here, we design an ultimately thin MS providing the same non-resonant electromagnetic response to an external BC and making the same effect to $B_1^{+}$ in the artifact region as a given DP optimized for abdominal imaging at 3 T. The unit cells of the proposed MS should create the same secondary $B_1^{+}$ field as same-sized fractions of the pad. Since non-resonant and small dielectric bodies support displacement currents (capacitive response), the same MS unit cells should do. With this aim, we introduced an MS comprised of a two-dimensional periodic structure of parallel-plate capacitors printed on two sides of a thin and flexible polyimide substrate and interconnected with thin copper strips. This capacitively-loaded grid was initially proposed for frequency-selective transmission of incident plane waves \cite{Anderson}. We show that with an adjusted capacitance of the parallel-plate elements, this structure slows down an external electromagnetic wave similar to a slab of ceramics with permittivity of 300 (optimal for abdominal imaging). Numerical simulations and \textit{in vivo} experiments confirm an equivalence between the effects of the pad and MS.

\section{Methods}

\subsection{Metasurface design and optimization}

\begin{figure*}
	\centering 
	\includegraphics[width=1\linewidth]{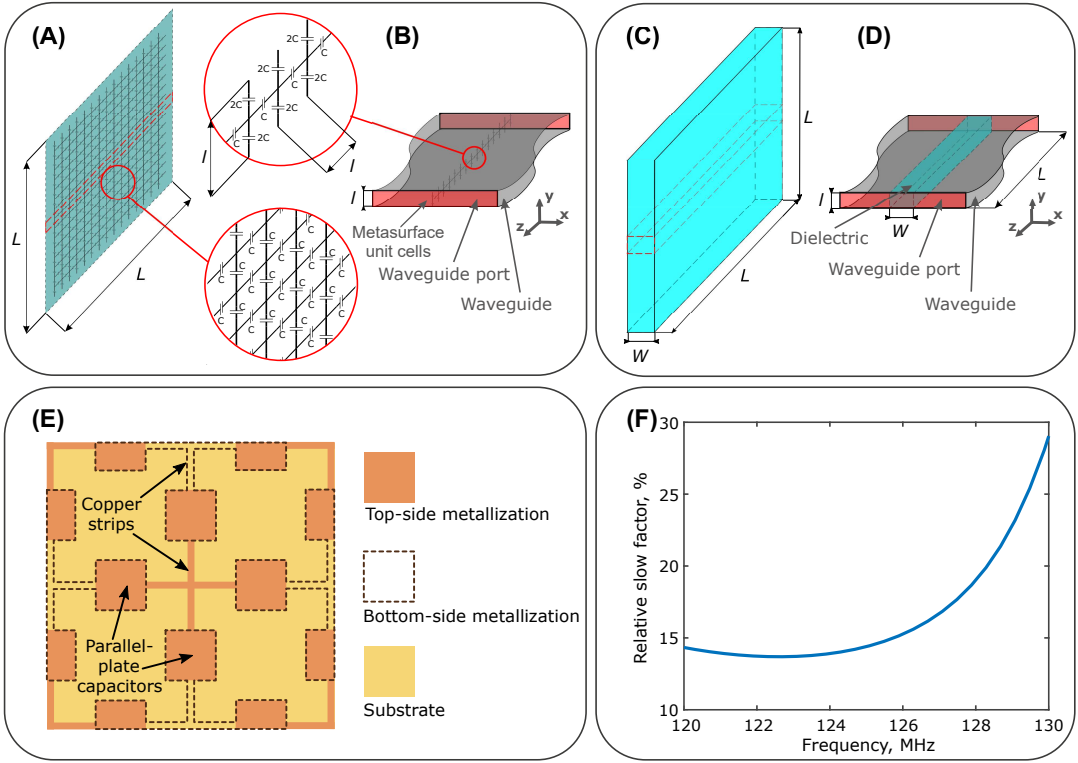}
	\caption{Metasurface (MS) modeling and comparison to a dielectric pad (DP): (A) schematic representation of the MS based on a capacitively-loaded grid (inset shows the topology of the strips loaded with capacitors); (B) one row of MS’ unit cells placed in a parallel-plate waveguide; (C) schematic representation of the DP; (D) fraction of a dielectric layer inside a parallel-plate waveguide; (E) Practical realization of the MS based on a dual-sided printed circuit board (four unit cells are shown); (F) Numerically calculated wave deceleration ratio (in percents) as a function of frequency.
	}
	\label{Fig1}  
\end{figure*}

The proposed MS constitutes a square grid of copper strips of width $w=0.1$ cm with periodicity $l=2$ cm printed on a thin dielectric substrate. Each strip is split in its center, and a capacitor with capacitance $C$ is connected to the split. Therefore a capacitively-loaded grid is formed \cite{Anderson}. The MS is schematically depicted in Fig.~\ref{Fig1}(A) with an inset showing the topology of the strips loaded with capacitors.

To behave similarly to a DP of the given permittivity, the unit cells of the MS should possess the same displacement currents as induced in the same-sized fractions of the dielectric later of the pad. For certainty, we considered a DP having the permittivity of approximately 300 found to be optimal for abdominal imaging with dimensions $L\times L = {28}\times{28}$ cm$^{2}$ and thickness $W=1.5$ cm, previously optimized for fetal imaging of 7-months pregnant woman \cite{van2019simulation}. The optimization goal for the proposed MS was to obtain the same displacement current in each unit cell as in the portion of the pad with the same in-plane dimensions $l\times l = {2}\times{2}$ cm$^{2}$. This condition should be met for an electromagnetic wave propagating along the MS. It is assumed that in the MS, all the displacement current is localized in the capacitors, which is valid for small cells in comparison to the wavelength. 

The value of the displacement current in the unit cell is controlled by capacitance $C$, which was determined from the numerical comparison of two models shown in Fig.~\ref{Fig1}(B,D). The first one is a single row of $n$=14 unit cells of the MS arranged in the $z$ direction and placed in a parallel-plate waveguide as shown in Fig.~\ref{Fig1}(B). The waveguide is composed of two metal plates placed parallel to $xz$ at a distance $l$ from each other. Since the $y$-directed strips in each unit cell are split and connected to the metal plates, two series-connected capacitors with double capacitance 2$C$ are modeled instead of one capacitor with capacitance $C$ as depicted in the inset of Fig.~\ref{Fig1}(B). Note that the dielectric substrate's presence has a negligible effect on the MS's properties and was omitted in the model of Fig.~\ref{Fig1}(B).
The second model is a fraction of the pad with the same overall dimensions $L\times l = {28}\times{2}$ cm$^{2}$ as the MS row and thickness $W=1.5$ cm, placed into the same parallel-plate waveguide (see Fig.~\ref{Fig1}(D)). The equivalence between both models implies that their capacitance per unit length in $z$-direction is the same, which can be observed in the numerical simulation as the same phase delay of a slow electromagnetic wave \cite{Slow} propagating between two waveguide ports. The above comparison shows the equivalence only to a wave propagating along the waveguide in $z$-direction with $y$-polarized electric field and $x$-polarized magnetic field. This scenario corresponds to the RF field characteristics at a pad's position at the periphery of a transmit BC when it is positioned parallel to the sagittal or coronal plane.

Simulations of both models shown in Fig.~\ref{Fig1}(B,D) were made using CST Microwave Studio commercial software (Frequency Domain Solver). The phase delay was determined as $\varphi=\text{arg}(S_{12})$, i.e., the phase of the complex transmission coefficient between the ports at the Larmor frequency of protons at 3 T ($f=123$ MHz). From the last quantity, an effective phase velocity $\upsilon_{\text{phase}}=(2\pi f\cdot nl)/\varphi$ of the wave propagating along the MS and DP can be estimated, which is much smaller as for free-space propagation in both cases. To compare the MS to the pad, we introduce the wave deceleration ratio (WDR) equal to $\text{WDR}=(\varphi_{\text{dielectric}}-\varphi_{\text{metasurface}})/{\varphi_{\text{dielectric}}}$. The best correspondence between the MS and DP is achieved at WDR of 0. 

We have chosen the MS period of 2 cm based on the reasons explained in the discussion section. For $l=2$ cm and $w=0.1$ cm, we found the capacitance $C=40$ pF for equivalence to the pad. In the practical realization of the MS, the capacitors were implemented as paired square parallel copper plates with dimensions of $d\times d = 5.8 \times 5.8$ cm$^{2}$ printed on the opposite sides of a substrate with thickness $t=25$~$\mu$m with relative permittivity $\varepsilon_s=3.4$ and dielectric loss tangent of 0.002. The capacitance was estimated using the formula of a parallel-plate capacitor: $C=\varepsilon_s\varepsilon_0d^2/t$ with $\varepsilon_0=8.85$ pF/m.
In the MS layout, the closest strips are also located at different sides of the substrate for properly connecting the adjacent capacitors, as shown in Fig.~\ref{Fig1}(E). The resulting WDR as a function of frequency is shown in Fig.~\ref{Fig1}(F), showing that the optimized MS provides the same phase delay as the pad with the accuracy of 13.5\% at 123 MHz.

\subsection{RF simulations}

The optimized pad was numerically characterized when placed into the RF field of a two-port high-pass BC tuned to 123 MHz in the presence of several voxel models of the human body. The $B_1^{+}$ field and local SAR patterns inside the body were calculated using CST Microwave Studio commercial software (Time Domain Solver). The transmit coil had a diameter of 70 cm, a length of 49 cm, and a shield diameter of 76 cm. These dimensions correspond to the BC on a standard-bore 3 T system. A set of three voxel body models from the CST Voxel Family was used in the simulations, including Emma (26-years-old female with BMI of 28.0 kg/m$^{2}$), Gustav (38-years-old old male with BMI of 22.3 kg/m$^{2}$), and Hugo (38-years-old male with BMI of 31.8 kg/m$^{2}$).

A high-permittivity DP with dimensions of ${28}\times{28}\times{1.5}$ cm$^{3}$, relative permittivity of 300, and conductivity of 0.4 S/m was used. This permittivity value was previously found to be optimal for abdominal imaging, and the dimensions were previously selected for fetal imaging of a 7-months pregnant woman~\cite{van2019simulation}. Simultaneously, the conductivity was found to have a minor effect on the pad's performance inside a BC. 

The MS was designed to have the same in-plane dimensions as the DP and was located in simulations at the same distance of 0.5 cm from the abdomen surface. The MS had the same shape of unit cells as described in the previous subsection for the practical realization with optimized square parallel-plate printed elements.

The $B_1^{+}$ field and local Specific Absorption Rate (SAR) were calculated assuming 1 W of accepted power and two perfectly matched BC' ports driven in quadrature. SAR was averaged over 10 g of body tissues. To analyze the effect of the pad or MS on $B_1^{+}$ inside the human body models, the coefficient of variation Cv was used. Cv (measured in percents) was calculated as the standard deviation of the $B_1^{+}$ magnitude divided by its mean value calculated across the given ROI in the abdominal region and multiplied by 100. This coefficient was chosen as $B_1^{+}$ field homogeneity figure of merit as proposed in \cite{de2012increasing}.

\subsection{Experimental study}

\begin{figure*}[ht]
	\centering 
	\includegraphics[width=0.7\linewidth]{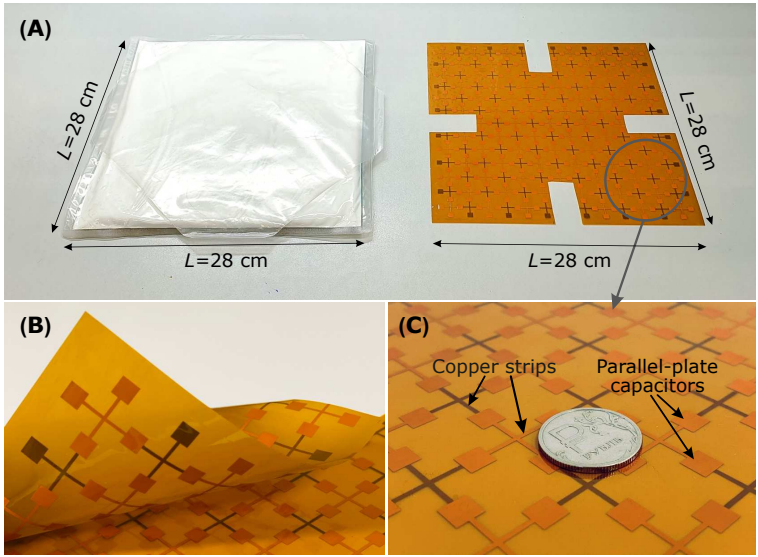}
	\caption{To the experimental comparison: (A) manufactured dielectric pad (on the left) and the metasurface (on the right) used in \textit{in vivo} tests. (B) Demonstration of the metasurface flexibility. (C) A close-up look at the metasurface showing its periodic copper pattern elements compared to a coin ($\approx$2 cm in diameter).
	}
	\label{Fig3}  
\end{figure*}

The manufactured high-permittivity DP shown in Fig.~\ref{Fig3}(A) had the same dimensions and properties as in the simulations. It was based on a polyethylene package filled with a powder of BaTiO${_3}$ ceramics mixed with heavy water. The weight of this pad was 3290 g. 

The MS shown in  Fig.~\ref{Fig3}(A,B,C) was manufactured as a flexible PCB with dual-layer copper metallization and a protective mask. The MS corresponds to the simulation model except for four rectangular cuts made on the edges. The cuts were made to avoid undesirable Fabry-Perot-type resonances at 123  MHz, as further discussed. The substrate material was DuPont Pyralyx AP8515R with a polyimide thickness of 25 $\mu$m, and a copper thickness of 18 $\mu$m, which made the MS easily flexible (Fig.~\ref{Fig3}(B)), as well as ultralight (the weight of the MS is equal to tens of grams). A close-up look at the MS's inner structure can be seen in Fig.~\ref{Fig3}(C). The semi-transparent substrate allows one to see copper strips printed on both sides: light-brown strips are on the top, and dark-brown ones are on the substrate's bottom. Each pair of neighboring strips from different sides is connected through a capacitance of two square plates printed one against the other.

\textit{In vivo} scanning was performed using Siemens Magnetom Trio A Tim 3 T whole-body MRI scanner (Siemens, Germany) with a bore diameter of 60 cm at the Almazov National Medical Research Center. An MRI sequence used for imaging was a T2-weighted HASTE sequence with acquisition parameters:  FA=$90^{\circ}$, TR/TE=$2000/97$ ms, acquisition matrix=$256\times179$, FOV=$306\times350$ mm$^{2}$. The receive coil used for the acquisition was either a Body Matrix Coil (local parallel receive coil with six channels) combined with two channels of Spine Matrix Coil or a BC (the same coil as in transmit). 
During \textit{in vivo} comparison of the pad and MS, three volunteers with different BMIs were tested: Volunteer \#1 with BMI of 33.5 kg/m$^{2}$, Volunteer \#2 with BMI of 28.6 kg/m$^{2}$, Volunteer \#3  with BMI of 27.9 kg/m$^{2}$. The scans were obtained under authorization by the local ethics committee (decision document dated July 20, 2020), and informed consent was obtained from each volunteer before the study.

\section{Results}

\subsection{RF simulations}

\begin{figure*}[ht]
	\centering 
	\includegraphics[width=1\linewidth]{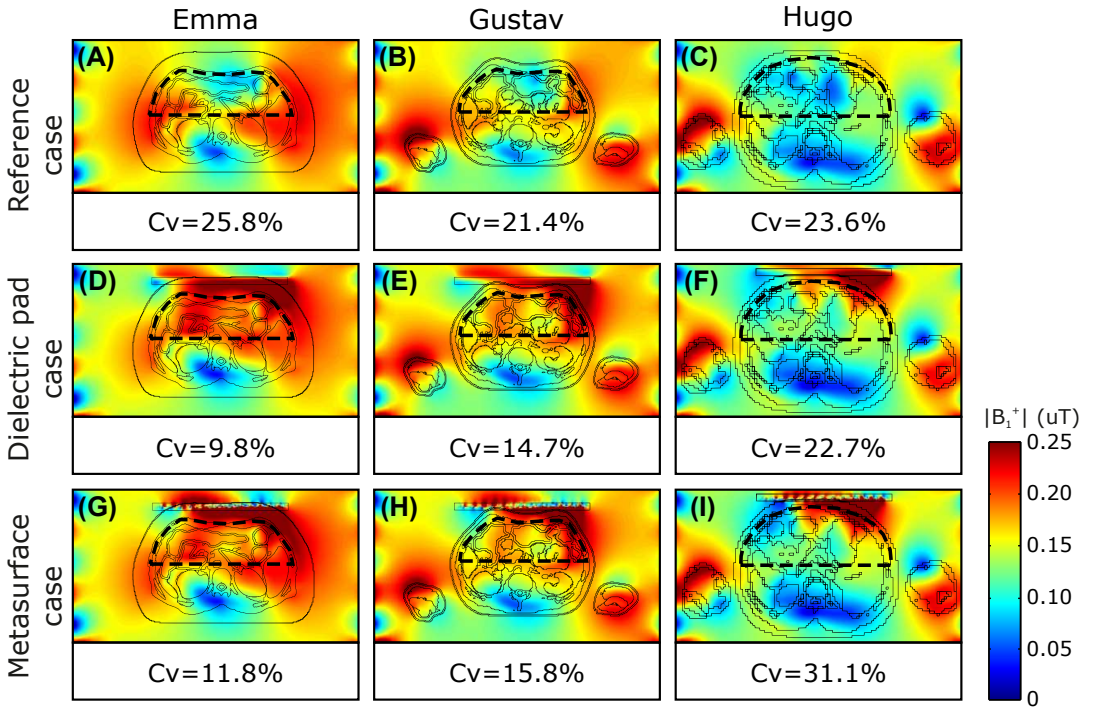}
	\caption{The numerically calculated magnitude of $B_1^{+}$ for three voxel models for the cases: (A,B,C) reference case with voxel model only; (D,E,F) with the dielectric pad attached to the top of the abdomen; (G,H,I) with the metasurface attached to the top of the abdomen.
	}
	\label{Fig4}  
\end{figure*}

Fig.~\ref{Fig4} shows simulated $B_1^{+}$ maps for the three different voxel models (Emma, Gustav, Hugo) in three cases: (A,B,C) reference case with the BC without any DP or MS; (D,E,F) with the DP or (G,H,I) with the MS attached on the top of the abdomen. The figure of merit was calculated as described in Methods and is denoted as Cv in each calculated $B_1^{+}$ map. The ROI used to calculate Cv for each case is shown by a dashed line. 

As can be seen, all three voxel models have inhomogeneous $B_1^{+}$ field created by the coil with a pronounced interference minimum in the central part of the ROI. In the reference case, the inhomogeneity defined by Cv is similar across all three models. However, the minimum is more visible and is characterized by a lower field level for bigger BMI values. For two voxel models (Emma and Gustav), the DP and MS decrease Cv and improve homogeneity. Note that the lowest Cv with the best homogeneity is achieved for Emma. For the Hugo voxel model, the MS increases the $B_1^{+}$ field in the nearby area but decreases the overall homogeneity Cv in the ROI. At the same time, the DP offers only a minor homogeneity improvement.

\begin{figure*}[ht]
	\centering 
	\includegraphics[width=1\linewidth]{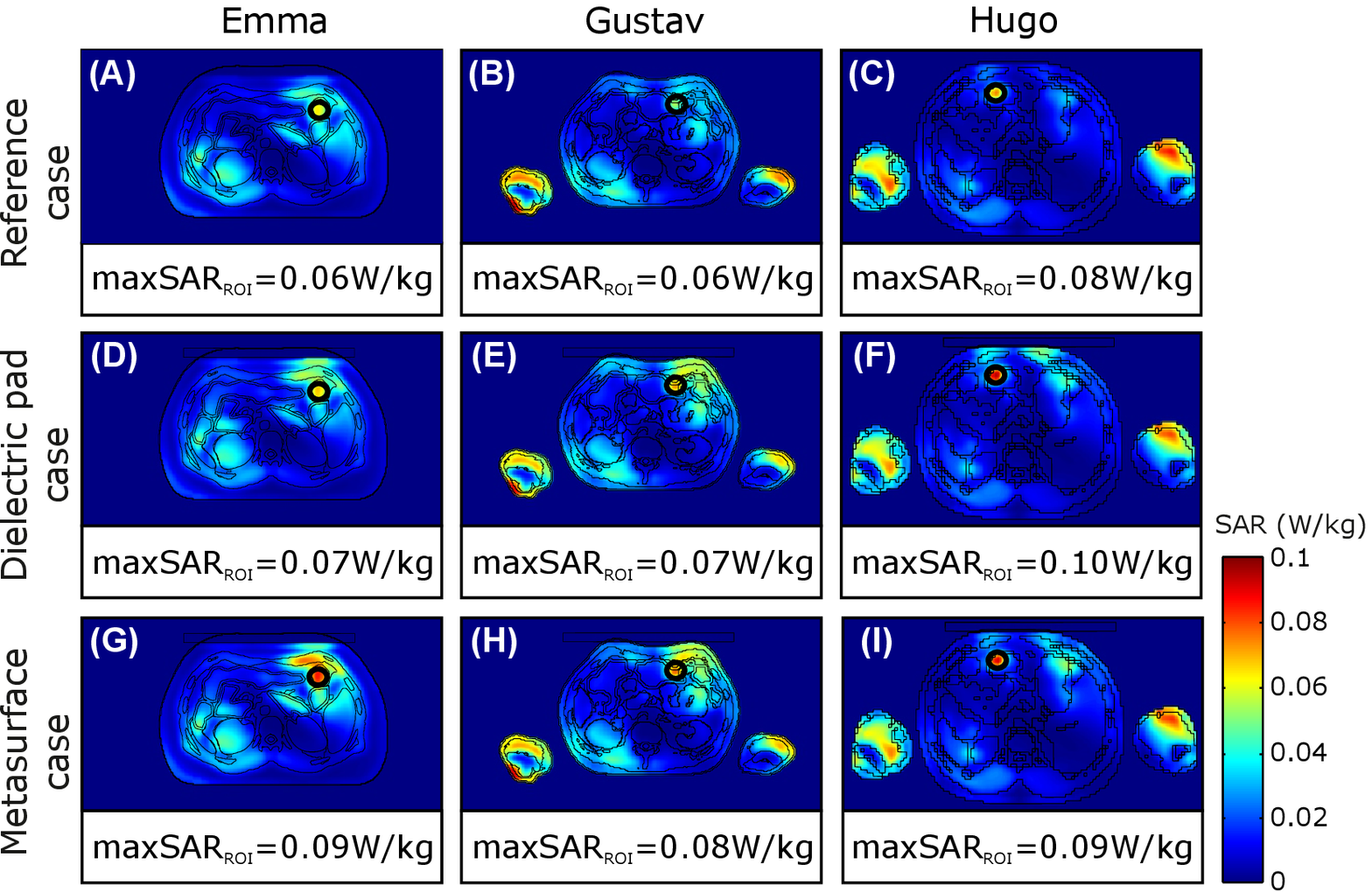}
	\caption{
	Local SAR averaged over 10 g of tissue of three voxel models for three cases: (A,B,C) reference case with voxel model only; (D,E,F) with the dielectric pad was attached on the top of the abdomen; (G,H,I) with metasurface attached on the top of the abdomen.
	}
	\label{Fig5}  
\end{figure*}

Fig.~\ref{Fig5} shows numerically calculated local SAR corresponding to the same three cases and three voxel models. The patterns are shown in the same cross-sections as $B_1^{+}$ in Fig.~\ref{Fig4}. The peak SAR value increases for all three voxel models when the DP or MS is inserted, while the maximum location does not change.


\subsection{Experimental study}

\begin{figure*}
	\centering 
	\includegraphics[width=0.9\linewidth]{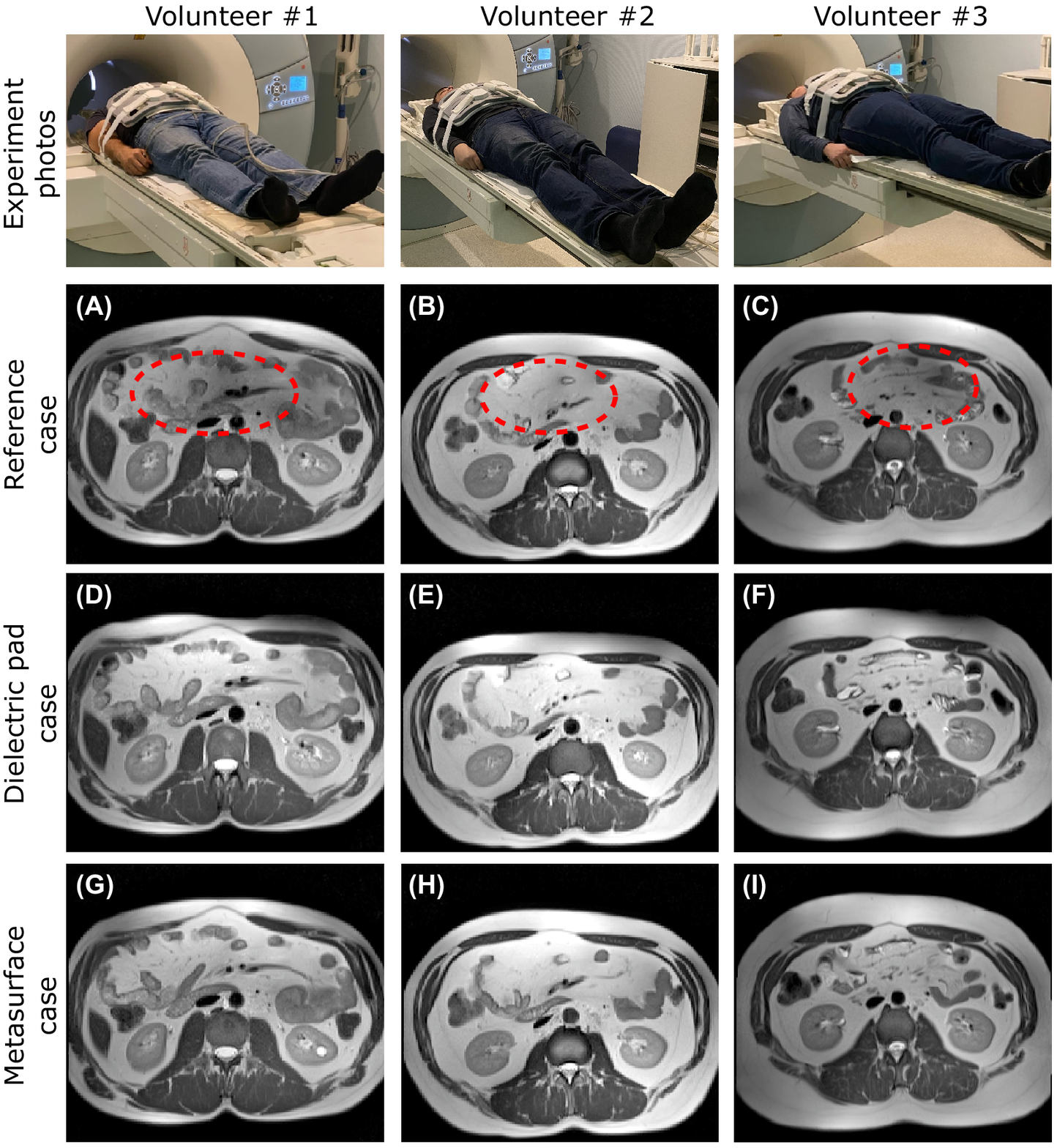}
	\caption{MR images of three healthy volunteers obtained with Body Matrix Coil used as a receive coil. Three cases were considered: (A,B,C) reference case with voxel model only; (D,E,F) with the dielectric pad was attached on the top of the abdomen; (G,H,I) with the metasurface attached on the top of the abdomen.
	}
	\label{Fig6}  
\end{figure*}

\begin{figure*}[ht]
	\centering 
	\includegraphics[width=0.9\linewidth]{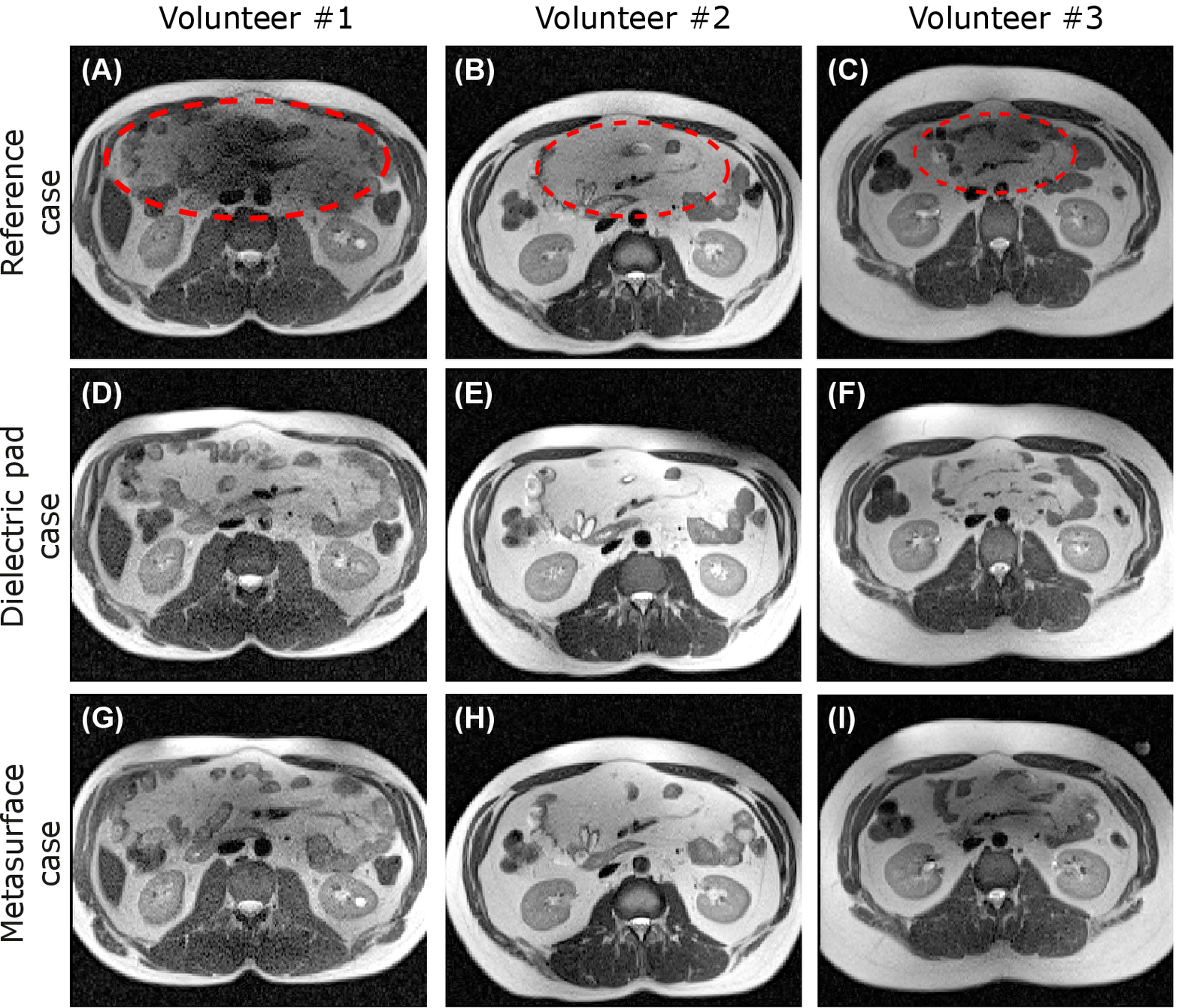}
	\caption{MR images of three healthy volunteers obtained with volumetric body coil used as a receive coil, for three cases: (A,B,C) reference case with voxel model only; (D,E,F) with the dielectric pad was attached on the top of the abdomen; (G,H,I) with the metasurface attached on the top of the abdomen.
	}
	\label{Fig7}  
\end{figure*}

Experimentally obtained MR images in the transverse plane of the abdominal cavity of three healthy volunteers with different BMI are shown in Fig.~\ref{Fig6} and Fig.~\ref{Fig7}. For the images shown in Fig.~\ref{Fig6}, the BC was used for transmitting, and a local Body Matrix Coil was used for receive. For the results presented in Fig.~\ref{Fig7}, we used the BC for both transmit and receive. In both experiments, we compared three cases: reference case with no DP or MS used; DP case, where the DP was attached on the top of an abdomen; MS case, where the MS was attached on the top of an abdomen. During the scans, the DP was placed directly on each volunteer's body without any spacers. When a local receive coil was used (see photographs in Fig.~\ref{Fig6}), the pad was placed between the coil and the body. The MS was covered from both sides with a 0.5-cm-thick foam spacer to avoid imaging artifacts caused by the periodicity of the secondary local RF field created by the MS unit cells. The bottom-side spacer was placed directly to the body. When a local receive coil was used, the coil was placed over the top-side spacer of the MS.  

In Fig.~\ref{Fig6}(A,B,C), a dark region with a low signal (marked by the red dashed line) can be seen in the top area of the abdomen for all three volunteers, being most prevalent for Volunteer \#1. With the DP (D,E,F), the dark region fades, and the signal intensity for each tissue in that area becomes more in line with the rest corresponding tissues in other areas of the abdomen. As can be seen from Fig.~\ref{Fig6}(G,H,I), with the MS, the dark region fades as well, and the images become visually similar to ones obtained with the pad. 

In Fig.~\ref{Fig7}(A,B,C), one can observe the artifact more clearly: the dark region is even larger than in the corresponding images of Fig.~\ref{Fig6}. The artifact shape is again different for three volunteers: the dark void is the most harmful for the image of Volunteer \#1, leading to an almost total signal loss.
In Fig.~\ref{Fig7}(D,E,F) and Fig.~\ref{Fig7}(G,H,I), one can see that both the pad and MS remove the dark region. Their effect is almost the same for Volunteer \#1, while for Volunteers \#2 and \#3, the pad provides slightly better homogeneity improvement.

Fig.~S1 in the supplementary data shows three different transverse slices of Volunteer \#1 acquired while using Body Matrix Coil for receive. One can observe that the dark region can be seen on all the slices, and both the DP and MS similarly remove the artifact on each of them. Supplementary Fig.~S2 shows the same three slices but scanned using the BC for receive. The result is the same as with Body Matrix Coil, but the dark area is visually more easily spotted. 

\section{Discussion and Conclusion}

The proposed MS was designed to have the same effect on the BC's transmit field as a DP, i.e., a flat dielectric layer made of a high-permittivity dielectric. It means that when excited by the transmit coil, each unit cell of the MS should support the same displacement current and create the same secondary $B_1^{+}$ field as the pad's fraction of the same dimensions. As was previously shown, the pad can create a sufficient secondary $B_1^{+}$ field level to remove the interference medium inside the human body's abdominal region when having the permittivity of about 300 and a thickness of 1.5 cm. Such a DP can only be constructed using quite expensive ceramic materials making the slab heavy. Indeed, a DP with several kilograms is uncomfortable for the patient to wear on the stomach during the scan.

On the other hand, the same induced displacement current per unit length of the pad can also be effectively induced in ultra-thin and flexible MS periodic unit cells. With this aim, we realized the unit cells as printed copper strips split and connected via parallel-plate printed capacitors. As all the metal parts are printed on two sides of a 25-$\mu$m-thick dielectric substrate, the entire MS is cheap to manufacture, flexible, and very light (less than 100 g together with a foam spacer). This miniaturization requires compact capacitors in each unit cell instead of a distributed ceramic material of the pad. The sufficiently high level of the unit-cell capacitance was achieved thanks to a small separation between printed parallel plates equal to the substrate thickness.

The parallel-plate capacitors' capacitance was optimized to have the same properties as the DP for abdominal imaging at 3 T with permittivity of 300. To adjust the capacitance, making the MS equivalent to the pad, we proposed a parallel-plate waveguide model analyzed numerically. In this model, the wave propagating along one row of MS unit cells or a break fraction of the pad with the same in-plane dimensions having an electric field polarized in parallel the MS/pad should experience the same phase delay. In other words, both systems behave as slow-wave transmission lines \cite{Slow,gorur1994novel, yang1998novel} with the same phase velocity at the Larmor frequency. The MS and DP excitation inside the waveguide is similar to real characteristics of the BC transmit field at its periphery. Indeed, the wave propagation direction and polarization correspond to a wave traveling around a BC's circumference when forming a fundamental CP mode. Advantageously, the proposed waveguide model is much simpler to optimize numerically than a full MS inside the coil. Once the MS cell's capacitance is optimized in the waveguide, the full-sized two-dimensional MS behaves similarly to the DP of the same in-plane dimensions. 

Despite the collective effect of the MS unit cells on $B_1^{+}$ field was the same, which was confirmed by full-wave numerical simulations (Fig.~\ref{Fig4}, Fig.~\ref{Fig5}) and imaging tests (Fig.~\ref{Fig6}, Fig.~\ref{Fig7}), there are some fundamental differences between a continuous ceramic slab and a periodic slow-wave structure that need to be considered when designing any new MS. The first significant limitation is that the secondary $B_1^{+}$ field of the MS compares to one of the DP only at distances from its plane larger than 1/4-1/2 of the period. Otherwise, the discreteness of the secondary $B_1^{+}$ may affect the image (the case illustrated in Fig.~S3 with a spatially-periodic signal modulation in the vicinity of the MS). Thus, the MS with a 2-cm-period should be separated from the body by at least 0.5 cm-thick foam spacer, while the DP can be placed onto the body without any gap. The second limitation consists in the limited slow-wave behavior. DPs reported in the literature usually have a relative permittivity in the range from 100 to 300. The lower the Larmor frequency and the static field $B_0$, the larger the required pad permittivity. However, a periodic MS can provide phase delays only smaller than $180^{\circ}$ per one period due to Bragg's frequency limit \cite{Slow}. For the chosen period of 2 cm, the maximum achievable phase delay corresponds to the effective permittivity limit of 3600. Generally, to reach slower wave propagation and approximate larger permittivity values and larger thicknesses of the pad, it is reasonable to keep the period as small as possible. On the other hand, as the period decreases, the unit cell's required capacitance becomes hard to implement using parallel plates printed on a substrate. In our case, 2 cm allowed us to accommodate parallel-plate capacitors with the required capacitance inside the unit cells using only one PCB layer. If a pad with higher permittivity and/or thickness is approximated, the capacitance should be increased while keeping the same period. With this aim, lumped surface-mount-device capacitors or multi-layer PCBs could be used.

Another critical issue of the proposed MS is the possible excitation of surface resonant modes (standing waves) inside the periodic structure, which needs to be avoided. As follows from the comparison with the waveguide model, both the high-permittivity slab and MS can guide slow-waves, which can, in principle, experience multiple reflections from the structure's edges and produce standing-wave mode resonances. While this edge effect in DPs is suppressed due to high losses, standing waves' excitation in the low-loss MS may result in high deviation from the expected effect on $B_1^{+}$. Similar edge effects were previously studied in arrays of capacitively coupled strips in \cite{7857715}. In the practical MS, all standing-wave resonances' spectral positions strongly depend on the unit cells' total amount. After manufacturing the PCB, the standing-wave resonances were controlled by measurements and made different from the Larmor frequency by slightly modifying the structure. We have cut from an initially square pattern two unit cells from each edge as shown in Fig.~\ref{Fig3}(A) to ensure that no narrow-band resonance is observable around 123 MHz. This fact was detected by estimating a surface-loop probe's reflection response placed horizontally over the MS center. The scheme of the setup and example curves of the frequency response before and after modification are shown in supplementary Fig.~S4. When working accidentally at the frequency of one of the standing-wave resonances, the MS turns to a focusing resonant surface coil and does not operate as a non-resonant pad.

In the previous study \cite{vorobyev2020artificial}, we have shown the feasibility of replacing a DP for brain imaging at 7 T with an artificial dielectric slab, limited to improving field homogeneity in the area that was not clinically valuable. In this work, we have successfully replaced a DP in its proper position against the body with an even thinner, two-dimensional, and flexible periodic structure, i.e., an MS. We designed the MS to operate similarly to a state-of-the-art DP for abdominal imaging at 3 T. Pads used in this application are especially heavy during large thicknesses and high density of ceramic powders. We have demonstrated in simulations and experimentally that the MS can remove the $B_1^{+}$ inhomogeneity artifact in the abdomen similar to the pad. Moreover, the proposed MS has shown the same stability of the field homogenization effect to a variation of BMI of the patient as the pad. No additional SAR hot-spot has been observed in the simulations.

The materials (polyimide films) required for manufacturing are widely available on the market and cheap. The manufacturing process itself, which comes down to printing a layout on a PCB, is a service that is also very widespread and inexpensive for mass production. The manufacturing cost of the considered prototype was around 100 USD. Unlike DPs, the MS weights incomparably less than one kilogram, and it will last longer in working order. The proposed topology and the explained design procedure can be adopted for replacing DPs with different thickness and permittivity values used in other imaging tasks at different static fields. We believe the proposed MS will become a reasonable alternative to DPs in clinical and research MRI.

\section*{Conflict of interest}
Authors declare no conflict of interests.

\bibliography{sample}

\newpage

\section*{Supplementary Material}

Improving B1 homogeneity in abdominal imaging at 3 T with light and compact metasurface
\\
\\
V. Vorobyev et al. 
\\
\\
The Supplementary Information includes 4 Supplementary Figures.
\\
\\
\setcounter{figure}{0}    
\begin{figure*}[ht]
	\centering 
	\includegraphics[width=0.9\linewidth]{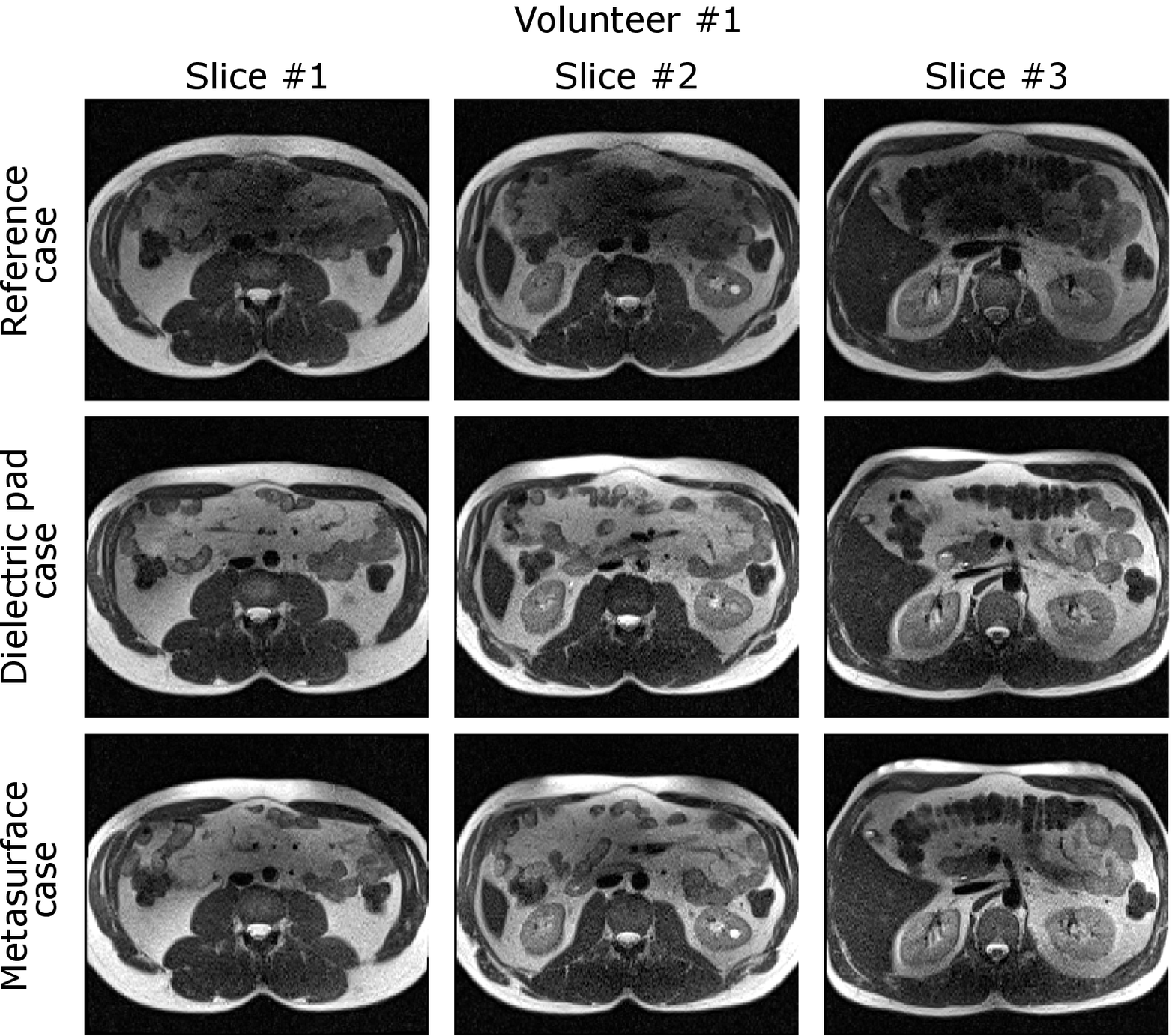}
	\renewcommand{\thefigure}{S\arabic{figure}}
	\caption{Experimentally measured MR scans of volunteer \#1 with Body Matrix Coil used as a receive coil, in three different axial slices for three cases: (1) reference case with no pad or metasurface used; (2) with the dielectric pad attached on the top of the abdomen; (3) with the metasurface attached on the top of the abdomen.
	}

\end{figure*}
\begin{figure*}[ht]
	\centering 
	\includegraphics[width=0.9\linewidth]{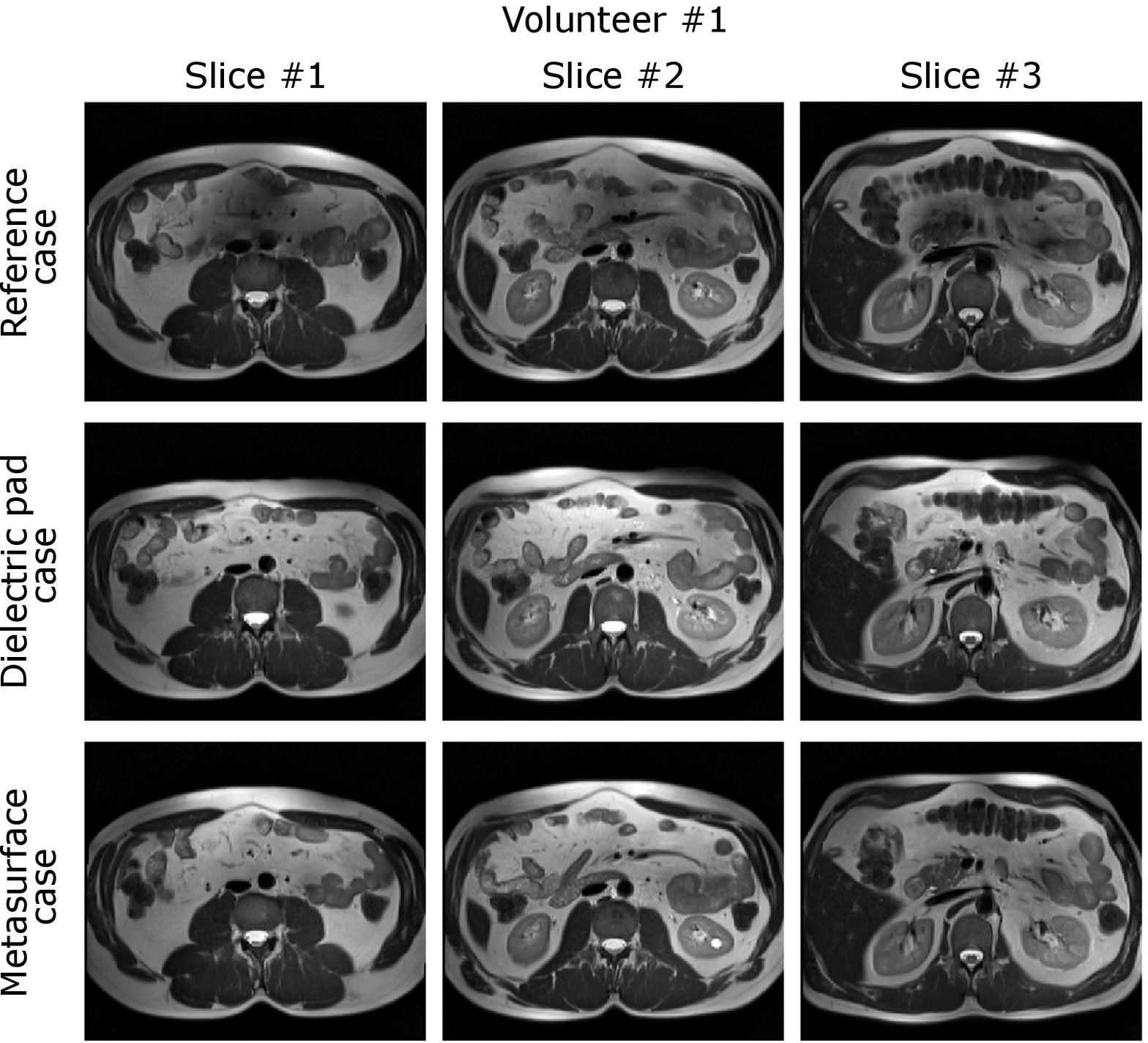}
	\renewcommand{\thefigure}{S\arabic{figure}}
	\caption{Experimentally measured MR scans of volunteer \#1 with volumetric body coil used in receive mode, in three different axial slices for three cases: (1) reference case with no pads or metasurface used; (2) with the dielectric pad attached on the top of the abdomen; (3) with the metasurface attached on the top of the abdomen.
	}

\end{figure*}
\begin{figure*}[hb]
	\centering 
	\includegraphics[width=0.9\linewidth]{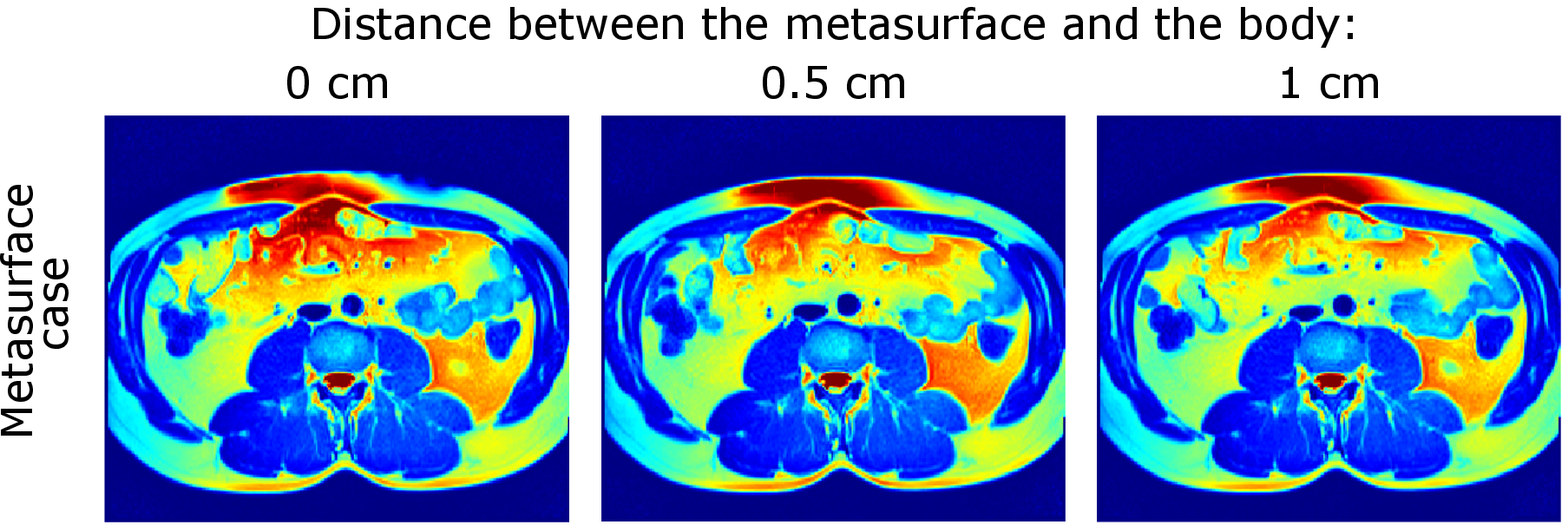}
	\renewcommand{\thefigure}{S\arabic{figure}}
	\caption{Experimentally measured MR scans of volunteer \#1 with Body Matrix Coil used as a receive coil and the metasurface. The figure shows how varying distance (0, 0.5 and 1 cm, respectively) between the metasurface and the body affects the resulting image.
	}

\end{figure*}
\begin{figure*}[t]
	\centering 
	\includegraphics[width=1\linewidth]{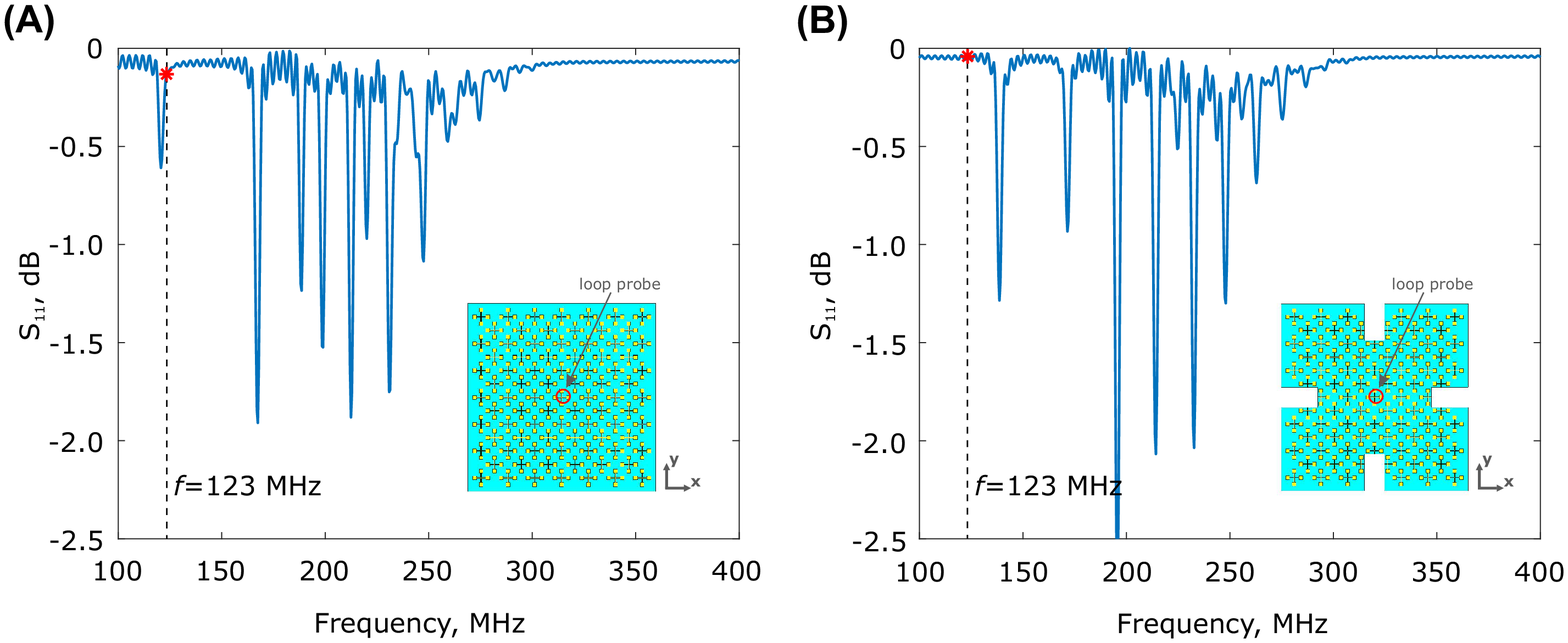}
	\renewcommand{\thefigure}{S\arabic{figure}}
	\caption{The reflection coefficient of the surface-loop probe was placed at a 1.5 cm distance from the metasurface center before (A) and after (B) metasurface modification.
	}

\end{figure*}

\end{document}